\documentstyle[aps,twocolumn,epsfig]{revtex}
\begin{document}
\title{Centrality  dependence  of  $\psi\prime$ over $J/\psi$ : a
signal of QGP}
\author{\bf A. K. Chaudhuri\cite{byline}}
\address{ Variable Energy Cyclotron Centre\\
1/AF,Bidhan Nagar, Kolkata - 700 064\\}
\maketitle
\begin{abstract}

If  nuclear  collisions  lead  to QGP formation then the ratio of
$\psi\prime$ over $J/\psi$ will remain  constant  with  $E_T$  as
both  $J/\psi$  and  $\psi\prime$  are  melted in the QGP. On the
other hand, if hot hadronic matter is produced,  the  ratio  will
continually   fall   with   $E_T$,  as  $\psi\prime$'s  are  more
suppressed in  hadronic  matter  than  the  $J/\psi$'s.  We  have
constructed  the  ratio  for  Pb+Pb  collisions  at  SPS from the
existing NA50 data. The ratio gives the indication of a  possible
QGP  formation  at SPS energy, but definite conclusion can not be
reached. We have also given the prediction for the ratio at  RHIC
energy. \end{abstract}


Lattice  QCD predicts that under certain conditions (sufficiently
high energy density and  temperature)  ordinary  hadronic  matter
(where  quarks  and  gluons  are  confined)  can  undergo a phase
transition to deconfined matter, commonly known  as  quark  gluon
plasma  (QGP).  $J/\psi$  suppression is recognized as one of the
promising  signal  of  the  deconfinement  transition.   Due   to
screening  of  color  force,  binding  of $c\bar{c}$ pairs into a
$J/\psi$ meson  will  be  hindered,  leading  to  the  so  called
$J/\psi$   suppression   in  heavy  ion  collisions  \cite{ma86}.
Experimental data indeed show suppression. However, all the  data
prior  to  NA50  Pb+Pb  are  well  explained  in terms of nuclear
absorption, also present in pA collisions \cite{vo99,ge99}.  NA50
collaboration  \cite{pbpb_jpsi}  observed  anomalous  suppression
(i.e. suppression beyond the normal nuclear  absorption)  in  158
GeV/c  Pb+Pb  collisions  \cite{pbpb_jpsi}. The ratio of $J/\psi$
yield to that of Drell-Yan pairs decreases faster with  $E_T$  in
the most central collisions than in the less central ones. It has
been  suggested that the resulting pattern can be understood in a
deconfinement  scenario  in  terms  of  successive   melting   of
charmonium   bound   states  \cite{pbpb_jpsi}.  However,  it  was
realized later that the data could be well explained in a variety
of models \cite{bl00,ca00,ch01,ch02,ch02b}, with or  without  QGP
formation. In ref\cite{ch02} it was also shown that the predicted
$J/\psi$  suppression  at  RHIC in a QCD based nuclear absorption
model, agree well with predictions obtained in a QGP based model.
It seems that, even at RHIC, deconfining phase  transition  could
not be detected from the $J/\psi$ suppression.

Not  only  $J/\psi$, but other states of charmoniums (e.g. $\chi$
and $\psi\prime$) are also suppressed in a QGP or  in  a  nuclear
matter.  In pA collisions, $\psi\prime$ suppression is similar to
$J/\psi$ suppression \cite{E777}. Recently in Quark matter  2002,
NA50    collaboration    confirmed   that   in   pA   collisions,
$\sigma^{J/\psi  N}_{abs}  \approx  \sigma^{\psi\prime  N}_{abs}$
\cite{pA_jpsi}.  They measured $J/\psi$ as well $\psi\prime$ from
Be, Al, Cu and W targets.  Parameterizing  the  production  cross
section  as  $\sigma^{pA}  \propto  A^\alpha$, NA50 collaboration
obtained, $\alpha_{J/\psi}
= 0.933 \pm 0.105$ and $\alpha_{\psi\prime} = 0.906  \pm  0.022$.
Nearly  identical  values  of  $\alpha$ for both the $J/\psi$ and
$\psi\prime$, in a Glauber type of model  of  nuclear  absorption
translate  into  similar  absorption cross section for them, i.e.
$\sigma^{J/\psi N}_{abs}  \approx  \sigma^{\psi\prime  N}_{abs}$,
contrary to the popular expectation that $\psi\prime$ being twice
as large in size than the $J/\psi$, $\sigma^{\psi\prime N}_{abs}$
will  be much larger than $\sigma^{J/\psi N}_{abs}$. The apparent
contradiction is resolved in the color octet model  \cite{octet}.
In  the  color  octet  model,  perturbatively produced $c\bar{c}$
pairs first neutralizes  its  color  by  combining  with  a  soft
collinear   gluon.   The  pre-resonance  $c\bar{c}g$  state  then
transforms  in  to  a  proper  charmonium  state,   $J/\psi$   or
$\psi\prime$.  In pA collisions, the nuclear medium sees only the
pre-resonance state. Equality of $\sigma^{J/\psi N}_{abs}$ and  $
\sigma^{\psi\prime N}_{abs}$ is then explained.

Unlike   in   pA  collisions,  in  AA  collisions,  $J/\psi$  and
$\psi\prime$   suppression   differs.   NA38/NA50   collaboration
measured   centrality  dependence  of  $J/\psi$  as  well  as  of
$\psi\prime$'s in S+U/Pb+Pb collisions  \cite{su_jpsi,pbpb_jpsi}.
Data  indicate that compared to $J/\psi$, $\psi\prime$'s are more
suppressed. For example, in S+U collisions,  from  peripheral  to
central  collisions  $J/\psi$'s  are  suppressed  by  a factor of
$\sim$ 1.3, while the $\psi\prime$'s are suppressed by  a  factor
of  $\sim$ 4. Similarly in Pb+Pb collisions, while $\psi\prime$'s
are suppressed by a factor of 8, $J/\psi$'s are suppressed  by  a
factor   of   three  only.  Thus  in  AA  collisions,  additional
suppression mechanism is operative for $\psi\prime$'s,  which  is
absent for the $J/\psi$'s in AA collisions.

One  of  the  source  for additional suppression could be the QGP
formation. If QGP formation  is  the  source  of  the  additional
suppression  of  $\psi\prime$'s,  why  the  effect is not seen in
$J/\psi$? Color screening studies shows that in a QGP, melting of
$J/\psi$  require  a   temperature   of   $1.2T_c$,   while   the
$\psi\prime$'s  are melted at $T_c$ only. Thus if QGP is produced
around $T_c$, its effect may be felt only on $\psi\prime$, not on
$J/\psi$'s. Also as the time scale of production of  $\psi\prime$
is  less  than  that of $J/\psi$, $\psi\prime$'s can better probe
the initial condition of the produced matter. Thus it is possible
that effect of QGP formation will be seen  only  in  $\psi\prime$
rather  than  in  $J/\psi$.  Hadronic comover's could also be the
source of additional suppression. In AA collisions a large number
of  secondaries  are  produced.  Absorption  cross   section   of
$\psi\prime$  in  comovers could be larger (due to larger radius)
than that of $J/\psi$'s,  leading  to  increased  suppression  of
$\psi\prime$.

In  the present letter we have analyzed the NA38/NA50 data on the
centrality dependence of $\psi\prime$ over  Drell-Yan  ratio,  in
S+U  and  in  Pb+Pb collisions. Analysis shows that absorption in
comovers or in QGP, both the scenario  could  explain  the  data.
Even  at RHIC energy, the ambiguity is not removed. It may not be
possible to detect deconfinement phase transition  from  $J/\psi$
or   $\psi\prime$   suppression.   Next  we  consider  centrality
dependence of the ratio of $\psi\prime$ over $J/\psi$. Gupta  and
Satz  \cite{gupta}  proposed  it  as a signal of QGP. The idea is
simple. If in the collision, QGP is formed  above  a  temperature
$1.2T_c$, then both the $J/\psi$ and $\psi\prime$ will be equally
suppressed  and  the  ratio  will  remain  constant  with  $E_T$.
Otherwise, the ratio will continually  decrease  with  $E_T$,  as
$\psi\prime$  are  more  suppressed  than $J/\psi$. The ratio has
been  considered  as  a   thermometer   for   the   deconfinement
temperature  also  \cite{sh97}. Though simple and quite old idea,
unfortunately, NA50 collaboration did not present  their  results
for  the said ratio for Pb+Pb collisions, which generated so much
interest about possible deconfinement phase transition. From  the
existing   data,   we   have  constructed  the  ratio  for  Pb+Pb
collisions. The centrality dependence of the ratio, though show a
tendency towards saturation,  it  is  not  possible  to  conclude
decisively  about  phase  transition.  Conclusive signal could be
obtained at RHIC energy.

In  the  QCD  based  nuclear  absorption model \cite{ch02,qiu98},
$J/\psi$ production is assumed to be  a  two  step  process,  (a)
formation of a $c\bar{c}$ pair, which is accurately calculable in
QCD  and  (b)  formation  of a $J/\psi$ meson from the $c\bar{c}$
pair, which is conveniently  parameterized.  The  $J/\psi$  cross
section  in  $AB$ collisions, at center of mass energy $\sqrt{s}$
is written as,

\begin{eqnarray} \sigma^{J/\psi} (s) &&
=K \sum_{a,b} \int dq^2 \left( \frac{\hat \sigma_{ab \rightarrow
cc}} {Q^2} \right) \int dx_F \phi_{a/A}(x_a,Q^2) \\ \nonumber
&&   \phi_{b/B}(x_b,Q^2)   \frac{x_a   x_b}{x_a   +  x_b}  \times
F_{c\bar{c} \rightarrow J/\psi} (q^2), \end{eqnarray}

\noindent  where  $\sum_{a,b}$  runs over all parton flavors, and
$Q^2 = q^2 +4 m_c^2$. The  $K$  factor  takes  into  account  the
higher  order corrections. The incoming parton momentum fractions
are fixed by kinematics and are $x_a
=(\sqrt{x^2_F+4Q^2/s}+x_F)/2$               and              $x_b
=(\sqrt{x^2_F+4Q^2/s}-x_F)/2$.
$\hat  \sigma_{ab \rightarrow c\bar{c}}$ are the subprocess cross
section and are given in \cite{be94}. $F_{c  \bar{c}  \rightarrow
J/\psi}(q^2)$  is  the  transition  probability that a $c\bar{c}$
pair with relative momentum square $q^2$ evolve into  a  physical
$J/\psi$ meson. It is parameterized as,

\begin{eqnarray} \label{4} F_{c \bar{c} \rightarrow J/\psi} (q^2)
= && N_{J/\psi} \theta(q^2) \theta({4m^\prime}^2 - 4 m_c^2 -q^2) \\
\nonumber
&&   (1  -  \frac{q^2}{{4m^\prime}^2  -  4  m_c^2  })^{\alpha_F}.
\end{eqnarray}

In  a  nucleon-nucleus/nucleus-nucleus  collision,  the  produced
$c\bar{c}$ pairs interact with nuclear medium before  they  exit.
It  is  argued  \cite{qiu98} that the interaction of a $c\bar{c}$
pair  with  nuclear  environment  increases  the  square  of  the
relative  momentum between the $c\bar{c}$ pair. As a result, some
of the $c\bar{c}$ pairs can gain enough relative square  momentum
to   cross   the   threshold  to  become  an  open  charm  meson.
Consequently,  the  cross  section  for  $J/\psi$  production  is
reduced  in comparison with nucleon-nucleon cross section. If the
$J/\psi$ meson travel a distance $L$,  $q^2$  in  the  transition
probability  is  replaced  to $q^2 \rightarrow q^2 +\varepsilon^2
L$, $\varepsilon^2$ being the relative square momentum  gain  per
unit  length.  Parameters  of the model ($\alpha_F$,$KN_{J/\psi}$
and $\varepsilon^2$) can be fixed from experimental data on total
$J/\psi$ cross section in pA/AA collisions. In Fig.1,  NA50  high
statistics  data \cite{pA_jpsi} are shown. Both the data sets are
well  explained  in   the   model   with   $\varepsilon^2$=0.1875
$GeV^2/fm$.  Nuclear suppression of $J/\psi$ and $\psi\prime$ are
due to  same  mechanism,  i.e.  gain  in  the  relative  4-square
momentum  of  the  $c\bar{c}$  pairs.  Naturally, $J/\psi$ and of
$\psi\prime$ shows similar A-dependence.

\begin{figure}[h]
\centerline{\psfig{figure=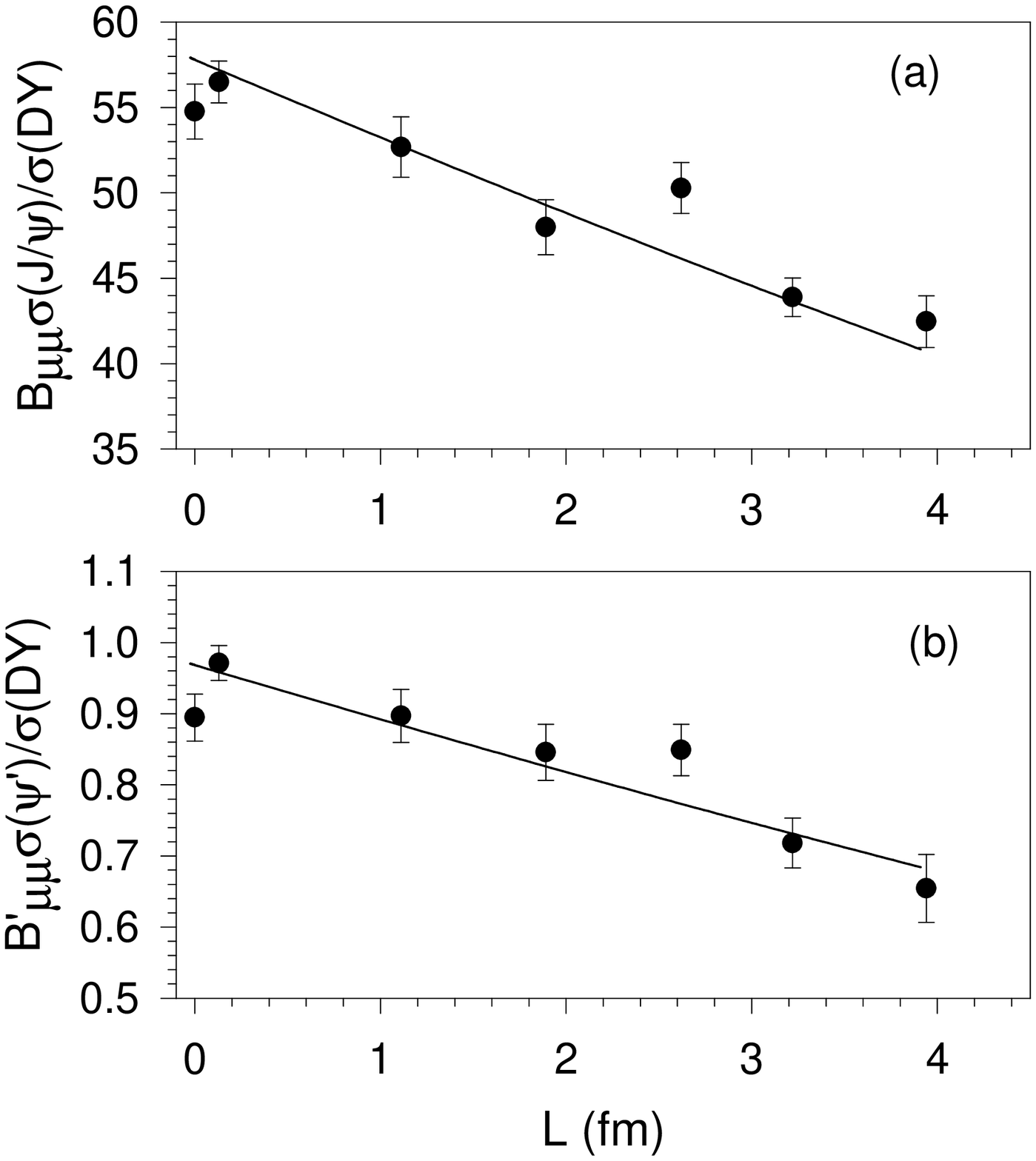,height=7cm,width=7cm}}
\vspace{-2.0cm}
\caption{(a)  The  experimental  ratio  of  total  $J/\psi$ cross
section and Drell-Yan cross sections in pp and pA collisions. The
solid  line  is  the  fit  obtained  in  the  QCD  based  nuclear
absorption model. (b) same as (a) for $\psi\prime$} \end{figure}

In  our  earlier  work  \cite{ch03}, we have shown that the model
could reproduce centrality dependence of $J/\psi$ over  Drell-Yan
ratio  in  S+U and Pb+Pb collisions. The $J/\psi$ or $\psi\prime$
cross sections at an impact parameter ${\bf b}$ as a function  of
$E_T$ can be written as,

\begin{equation}  \label{c1}
\frac{d^2\sigma^{J/\psi,\psi\prime }}{dE_Td^2b}      =
\sigma^{J/\psi,\psi\prime}_{NN}  \int  d^2s T_A(s) T_B({\bf b-s})
S(L({\bf b,s})) P(b,E_T) \end{equation}

\noindent  where  $T_{A,B}$  is  the  nuclear thickness function,
$T_{A,B}(b)=\int dz \rho(b,z)$. For the density, we have used the
three parameter Fermi distribution \cite{jvv},

\begin{equation}
\rho(r)=\frac{\rho_0(1+\omega\frac{r^2}{C^2})}{1+exp(-(r-C)/a)};
\hspace{5mm} \int \rho(r)d^3r=A \end{equation}

The parameters of the density distribution ($C$,$\omega$ and $a$)
are taken from \cite{jvv}.

$P(b,E_T)$ is the $E_T-b$ correlation function. We have used  the
Gaussian form for the $E_T-b$ correlation,

\begin{equation}\label{5}             P(b,E_T)            \propto
exp(-(E_T-qN_p(b))^2/2q^2aN_p(b)) \end{equation}

\noindent where $N_p(b)$ is the number of participant nucleons at
impact  parameter  b.  $a$  and  $q$  are  parameters  related to
dispersion and average transverse energy. For S+U  collisions  at
200   GeV/c,   the  parameters  are,  $a$=3.2  and  $q$=0.74  GeV
\cite{vo99}, and for 158 GeV/c Pb+Pb  collisions  the  parameters
are, $a$=1.27 and $q$=0.274 GeV \cite{bl00}.

In  Eq.\ref{c1}  $S(L)$  is the suppression factor due to passage
through a length L in nuclear environment. At an impact parameter
${\bf b}$ and at point  ${\bf  s}$,  the  transverse  density  is
calculated as,

\begin{equation}               n({\bf              b,s})=T_A({\bf
s})[1-e^{-\sigma_{NN}T_B({\bf b-s})}]  +  [A  \leftrightarrow  B]
\end{equation}

\noindent  and  the  length  $L({\bf  b,s})$ that the $J/\psi$ or
$\psi\prime$ meson will traverse is obtained as,

\begin{equation} L({\bf b,s})=n({\bf b,s})/2\rho_0 \end{equation}

The  Drell-Yan  pairs  do not suffer final state interactions and
the cross section at an impact parameter ${\bf b}$ as a  function
of $E_T$ could be written as,

\begin{equation}         \frac{d^2\sigma^{DY}}{dE_Td^2b}        =
\sigma^{DY}_{NN}  \int  d^2s  T_A(s)  T_B({\bf   b-s})   P(b,E_T)
\end{equation}

The  additional  suppression required for $\psi\prime$ may be due
to QGP formation or due to comover  interactions.  To  take  into
account the suppression due to QGP formation we assume that above
a  threshold  density,  $n_c$, all the $\psi\prime$ are dissolved
\cite{bl00}, and  introduce  the  additional  suppression  factor
$S_{QGP}$ in Eq.\ref{c1},

\begin{equation}
S_{QGP}({\bf b,s}) = \Theta(n_c-\frac{E_T}{<E_T>({\bf b})}n({\bf b,s})),
\end{equation}

The  additional suppression factor in the comover scenario can be
written as \cite{ga96},

\begin{equation}     S_{co}({\bf    b,s})=exp(-\sigma_{co}v_{rel}
n_0{(\bf b,s)}\tau_0 \ln(R_T/v_{rel}\tau_0)) \end{equation}

In  the  above  equation, $\sigma_{co}$ is the comover absorption
cross section for  the  $\psi\prime$'s,  $v_{rel}$  =0.6  is  the
relative  velocity  of  $\psi\prime$ with respect to comovers and
$\tau_0$=2 fm, is the time beyond which the comover  interactions
starts. $R_T$ is the transverse radius of the system and $n_0$ is
the  initial  comover  density.  To  account for the variation of
density  with  $E_T$,  we  take  $n_0=<n_0>E_T/<E_T>(b=0)$,  with
$<n_0>$=0.8  $fm^{-3}$ \cite{ga96}. The only quantity to be fixed
is the $\sigma_{co}$, which we obtained directly from fitting the
S+U data.

The        other        unknown       quantity       is       the
$N^{J/\psi/\psi\prime}=B_{\mu\mu}\frac{\sigma(J/\psi/\psi\prime)_{NN}}
{\sigma(DY)_{NN}}$. Experimentally it is known  for  450  GeV  pp
collisions      \cite{pA_jpsi}.      Craigie     parameterization
\cite{craigie} of DY cross sections could be used to  obtain  its
value  at  other  energies.  For  200  GeV/c  S+U collisions, the
extrapolated     values     are,      $N^{J/\psi}$=32-42      and
$N^{\psi\prime}$=0.53-0.68,  for  the  DY  invariant  mass in the
ranges of 2.1-3.1 GeV. We obtain the values of  $N^{J/\psi}$  and
$N^{\psi\prime}$ from a constraint fit to the NA38 data such that
they  are within the range of extrapolated values. For $J/\psi$'s
in Pb+Pb collisions, we rescale the value  by  the  factor  1.051
\cite{pA_jpsi}. For $\psi\prime$'s, we use the same value.

In  Fig.2a,  we  have  shown  the  NA38  data  on  the centrality
dependence of $J/\psi$ over Drell-Yan ratio, for  200  GeV/c  S+U
collisions.  The  solid line is the fit to the data obtained with
$B_{\mu\mu}\sigma^{J/\psi}_{NN}/\sigma^{DY}_{NN}$=39.02. Data are
well explained. In Fig.2b, the latest NA50 data on the centrality
dependence of the ratio of $J/\psi$ over Drell-Yan are shown. The
solid line is the prediction in the QCD based nuclear  absorption
model,         with         the        normalising        factor,
$B_{\mu\mu}\sigma^{J/\psi}_{NN}/\sigma^{DY}_{NN}$=41.01.      The
latest  NA50  data  are also well explained in the model. We note
that there is no scope for additional suppression due to  comover
interaction  or  due to QGP formation. Thus centrality dependence
of $J/\psi$ suppression in S+U or  in  Pb+Pb  collisions  do  not
require  additional  suppression  due  to QGP formation or due to
comover interactions.

\begin{figure}[h]
\centerline{\psfig{figure=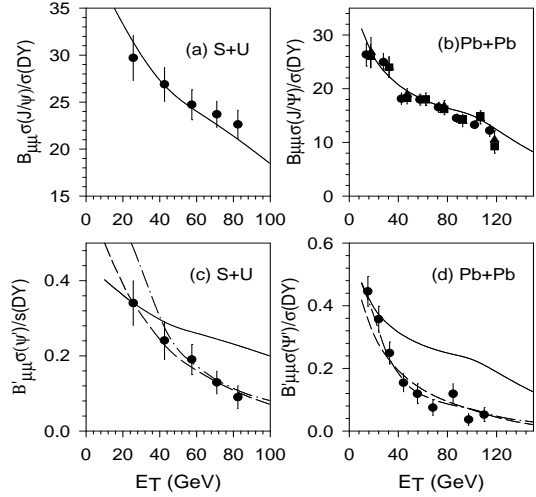,height=9cm,width=7cm}}
\vspace{-2.0cm}  \caption{(a) Experimental data on the centrality
dependence of $J/\psi$ over Drell-Yan ratio,  in  200  GeV/c  S+U
collisions.  The  solid line is the fit obtained to the data in
the QCD based nuclear absorption model. (b) same as (a) for Pb+Pb
collisions. (c)Experimental data on the centrality dependence  of
$\psi\prime$  over  Drell-Yan ratio in S+U collisions. The solid,
dashed and dash-dotted  lines  are  the  fit  to  the  data  with
nuclear,     nuclear+comover    and    nuclear+QGP    suppression
respectively. (d) same as (c) for Pb+Pb.} \end{figure}

The centrality dependence of $\psi\prime$ over Drell-Yan ratio on
the  other  hand require additional suppression. In Fig.2c and 2d
the       NA38/NA50        data        on        the        ratio
$B_{\mu\mu}\sigma(\psi\prime)/\sigma(DY)$  for  200 GeV/c S-U and
for 158 GeV/c Pb+Pb collisions are shown. The solid line  is  the
ratio  obtained  in  the  QCD  based  nuclear  absorption  model.
Irrespective                                                   of
$B_{\mu\mu}\frac{\sigma^{\psi\prime}_{NN}}{\sigma^{DY}_{NN}}$
value, the model clearly fails to explain  both  the  data  sets.
$\psi\prime$'s  are  not  sufficiently  suppressed  to agree with
experiment. As discussed in the beginning, additional suppression
could be either due to comovers  or  due  to  QGP  formation.  In
Fig.2c  and  d,  the  dashed  line  is  the  ratio  obtained with
nuclear+comover suppression, with $\sigma_{co}$=8  mb.  For  both
the         data         sets,         we        have        used
$B_{\mu\mu}\frac{\sigma^{\psi\prime}_{NN}}{\sigma^{DY}_{NN}}$=0.59,
obtained from fitting the NA38 S+U  data.  The  comover  scenario
fits  the  $E_T$  dependence  of $\psi\prime$ in S+U and in Pb+Pb
collisions.  For  Pb+Pb  collisions,  for  the  very   peripheral
collisions,  model  produces more suppression than in data. Since
we have fixed the comover  density  in  central  collisions,  the
simple ansatz may be inaccurate for peripheral collisions.

In  S+U  collisions,  centrality  dependence of $\psi\prime$ over
Drell-Yan ratio is not well explained if the nuclear  suppression
is   augmented   with  suppression  due  to  QGP  formation  (the
dash-dotted line). At low $E_T$ data are not explained.  Also  we
obtain a threshold density, $n_c$=1.8 $fm^{-2}$, which is too low
for  QGP  formation.  For  Pb+Pb  collisions  on  the  other hand
(Fig.2d), centrality dependence of  $\psi\prime$  over  Drell-Yan
ratio   are   rather   well   explained  with  nuclear  plus  QGP
suppression. In  Fig.2d,  the  dash  dotted  line  is  the  ratio
obtained  with  threshold  density, $n_c$=2.8 $fm^{-2}$. Data are
well explained throughout the $E_T$ range.  However,  as  nuclear
plus  comover  suppression  also  explain  the  data,  it  is not
possible to conclude positively about the formation of  QGP  from
the $E_T$ dependence of $\psi\prime$ suppression.

With  RHIC  being  operational,  it  is  interesting  to  predict
suppression at RHIC energy. At RHIC energy, the  so  called  hard
component,   which  is  proportional  to  the  number  of  binary
collisions, appear. Model dependent  calculations  indicate  that
the  hard component grows from 22\% to 37\% as the energy changes
from 56 GeV to 130 GeV \cite{kh01}. In our calculation,  we  have
used 37\% hard scattering component.

In  Fig.3,  we  have shown the predicted centrality dependence of
the $ \psi\prime$ over Drell-Yan ratio at RHIC energy  for  Au+Au
collisions.   The   solid   and   dashed   lines  corresponds  to
nuclear+comover and  nuclear+QGP  absorption  respectively.  They
agree  closely  with each other. $E_T$ dependence of $\psi\prime$
over Drell-Yan ratio at RHIC also could not  distinguish  between
the two scenarios.

\begin{figure}[h]
\centerline{\psfig{figure=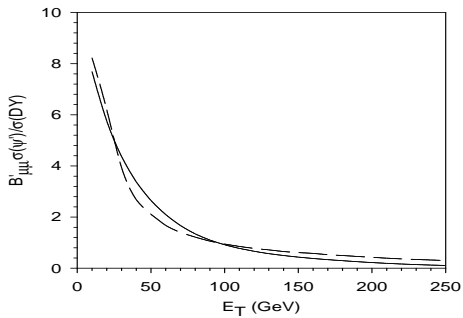,height=7cm,width=7cm}}
\vspace{-2.5cm}  \caption{Centrality  dependence  of $\psi\prime$
over Drell-Yan ratio at RHIC. The  solid  and  dashed  lines  are
obtained   with   nuclear+comover   and  nuclear+QGP  suppression
respectively. } \end{figure}

Next  we  consider the centrality dependence of $\psi\prime$ over
$J/\psi$. As told earlier, it has been proposed as  a  signal  of
the   QGP   formation.  The  proposal  follows  from  the  simple
observation that in a QGP both the $J/\psi$ and $\psi\prime$ will
be        melted.         Consequently,         the         ratio
$\sigma(\psi\prime)/\sigma(J/\psi)$  will  remain  constant  with
$E_T$. Otherwise, the ratio will continually fall with $E_T$,  as
$\psi\prime$ are more suppressed than $J/\psi$. In Fig.4, we have
tested  the  proposition.  NA50 collaboration did not present the
data. From the existing $J/\psi$ and $\psi\prime$  data  we  have
constructed  the  ratio. It is shown in Fig.4. The ratio, for S+U
collisions is also shown in Fig.4. For S+U collisions, the  ratio
fall   continuously   with  $E_T$.  QGP  is  not  formed  in  the
collisions. For the Pb+Pb collisions, the ratio falls with  $E_T$
till around 70 GeV and thereafter shows a tendency of saturation.
Data  do  not cover enough $E_T$ range for a definite conclusion.
In Fig.4, the solid and the dashed lines are the ratio for  Pb+Pb
collisions in the nuclear+comover and nuclear+QGP suppression. As
expected  both  of them fits the data. We note that even at large
$E_T$, difference between the two  model  calculations  is  small
(1-2\%).  Even  if  there  is  a  phase  transition,  it  will be
difficult to reach a definite conclusion.

The  situation  is much better at RHIC energy. Our prediction for
the ratio at RHIC is shown in Fig.4. The  dash-dot  line  is  the
prediction  obtained  with  nuclear+comover  absorption. It shows
continual fall  of  the  ratio.  In  contrast,  with  nuclear+QGP
suppression  (the  dash-dot-dot  line), the ratio remain constant
for $E_T >$ 70 GeV. The difference between the two predictions is
also large and easily detectable.

\begin{figure}[h]
\centerline{\psfig{figure=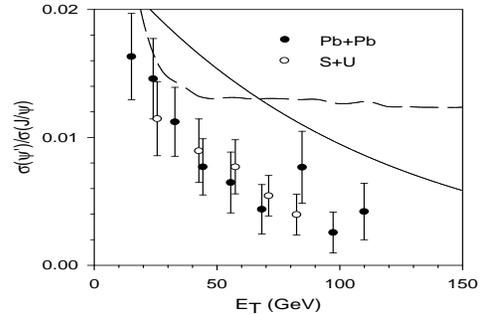,height=7cm,width=7cm}}
\vspace{-2.5cm}   \caption{   $E_T$   dependence  of  the  ratio,
$\psi\prime$ over $J/\psi$ in 200 GeV/c  S+U  and  in  158  GeV/c
Pb+Pb  collisions.  The  solid and dashed lines are obtained with
nuclear+comover and nuclear+QGP suppression. The predicted  ratio
at  RHIC  energy with nuclear+comover suppression is shown as the
dash-dot  line.  The  dash-dot-dot  line  is   the   ratio   with
nuclear+QGP suppression.} \end{figure}

To  conclude,  we  have  analyzed  the  centrality  dependence of
$J/\psi$  and  $\psi\prime$  suppression  in  S+U  and  in  Pb+Pb
collisions.  It  was shown that while the $J/\psi$ suppression is
well explained in the QCD based  nuclear  absorption  model,  the
model could not explain the centrality dependence of $\psi\prime$
suppression.   $\psi\prime$'s   require  additional  suppression,
either due to QGP formation or due  to  comovers,  two  scenarios
could  not be distinguished, even at RHIC. We then considered the
$E_T$ dependence of the ratio of $\psi\prime$ over $J/\psi$ as  a
signal  for  the  deconfining  phase transition. If QGP is formed
following a deconfinement phase transition, the ratio will remain
constant with $E_T$ in contrast to the  continuous  fall  of  the
ratio   in   case   of   no   such  formation.  The  experimental
$\sigma(\psi\prime)/\sigma(J/\psi)$ in Pb+Pb  collisions  is  not
conclusive.  However, at RHIC energy, the ratio could distinguish
between the comover and QGP suppression.

\end{document}